\documentclass[epj,referee]{svjour}
\usepackage{graphicx}
\usepackage{dcolumn}
\usepackage{bm}
\usepackage{amssymb}
\usepackage{amsmath}
\usepackage[usenames]{color}

\newcommand{\col}[1]{#1}

\begin{document}

\title{Effect of noise on generalized synchronization of chaos: theory and experiment}
\author{Olga~I.~Moskalenko\thanks{\emph{e-mail:} o.i.moskalenko@gmail.com}, Alexander~E.~Hramov, Alexey~A.~Koronovskii \and Alexey~A.~Ovchinnikov}
%
%
\institute{Faculty of Nonlinear Processes, Saratov State University,
Astrakhanskaya, 83, Saratov, 410012, Russia}
\date{Received: date / Revised version: date}
%
\abstract{The influence of noise on the generalized synchronization
regime in the chaotic systems with dissipative coupling is
considered. If attractors of the drive and response systems have an
\col{infinitely large} basin of attraction, generalized
synchronization is shown to possess a great stability \col{with
respect} to noise. The reasons of the revealed particularity are
explained by means of the modified system approach [A.~E.~Hramov,
A.~A.~Koronovskii, Phys. Rev. E. \textbf{71}, 067201 (2005)] and
confirmed by the results of numerical calculations and experimental
studies. The main results are illustrated using the examples of
unidirectionally coupled chaotic oscillators and discrete maps as
well as spatially extended dynamical systems. Different types of the
model noise are analyzed. Possible applications of the revealed
particularity are briefly discussed. }



\authorrunning{O.I.~Moskalenko et al.}
\maketitle

\sloppy

\section*{Introduction}
\label{sct:Introduction} Synchronization is one of the most relevant
directions of nonlinear dynamics attracting great attention of
modern
scientists~\cite{Pikovsky:2002_SynhroBook,Boccaletti:2002_ChaosSynchro}.
The interest to it is connected both with a large fundamental
significance of its investigation~\cite{Pikovsky:2002_SynhroBook}
and a wide practical applications, e.g. for the transmission of
information~\cite{Roy:Chaos_Nature2005,Jaeger:Science2004ChaosComm,Parlitz:1992_ChaoticCommunication,%
Cuomo:1993_ChaosCommunication,Kocarev:1995_ChaoticComunication,Peng:1996_GyperChaosSynchro,%
Eguia:2000,Fischer:2000_ChaosCommunication,%
Rulkov:2002_ChaoticCommunication,Yuan_SecureCommunication:2005,Li_ChaoticTrans:2006,Fradkov_ChaosTrans:2006,%
Cessac:TransSign2006,Rohde:EstSecCom2008,Materassi:TimeScalingSecure:2008,alkor:2010_SecureCommunicationUFNeng},
diagnostics of dynamics of some biological systems~\cite{Strogatz:1994_ChaosBook,Elson:1998_NeronSynchro,Porcher:2001_LEinMedicine,%
Glass:2001_SynchroBio,Pavlov:MultiscalityPhysA2002,Rosenblum:2004_SynchroBioSystems},
control of chaos in the microwave
systems~\cite{Ditto:1990_ControllingChaos,Ticos:2000_PlasmaDischarge,%
Rosa:2000_PlasmaDischarge,Hramov:2005_Chaos_BWO,dmitriev:074101},
etc.

Several types of the synchronous chaotic system behavior are
traditionally distinguished. They are
phase~\cite{Rosenblum:1996_PhaseSynchro,Pikovsky:2002_SynhroBook},
generalized~\cite{Rulkov:1995_GeneralSynchro,Kocarev:1996_GS},
lag~\cite{Rosenblum:1997_LagSynchro,Taherion:1999_LagSynchro},
complete~\cite{Pecora:1990_ChaosSynchro,Pecora:1991_ChaosSynchro},
time
scale~\cite{Hramov:2004_Chaos,Aeh:2005_TSS:PhysicaD,Hramov:2005_Chaos_BWO}
synchronization and others.

One of the most important problems connected with the study of the
chaotic systems is the influence of noise on their behavior
including the synchronous regime
arising~\cite{Stone:726,Kifer:StochMathJ1981,Ali:1997_NoiseSynchro,Toral:2001_NoiseSynchro,Anishchenko:2002_SynchroEng,%
Zhou:PRL2002,Zhou:2002_NoiseEnhancedPS,Kuznetsov:2003_WeakSynchroAndNoise,Zhou:PRE2003,%
Goldobin:2005_SynchroCommonNoise,Hramov:2006_PLA_NIS_GS,Guan:NoiseGS:2006}.
Noise is known to influence on system dynamics in different ways. In
particular, in case of complete synchronization of coupled chaotic
oscillators, noise may induce intermittent loss of synchronization
due to local instability of the synchronization
manifold~\cite{Heagy:CSNoise_PRE1995,Gauthier:PRL1996_CSNoise}. At
the same time, both periodic and chaotic non-coupled identical
dynamical systems subjected to a common noise may achieve complete
synchronization at a large enough
intensity~\cite{Fahy:1992_NoiseInfluence,Martian:1994_SynchroNoise,%
Jensen:1998NISPeriodic,Toral:2001_NoiseSynchro,Goldobin:2005_SynchroCommonNoise}.
Such phenomenon is called noise-induced synchronization regime. In
the case of phase synchronization of coupled oscillators noise can
induce phase slips in phase-locked periodic and chaotic
oscillators~\cite{Zhu:2001_transient,Hramov:2007_TypeIAndNoise}. On
the other hand, noise can play a constructive role at phase
synchronization enhancing the synchronous regime below the threshold
of phase
synchronization~\cite{Zhou:2002_NoiseEnhancedPS,Hramov:ZeroLE_PRE2008}.
Nevertheless, for almost all types of chaotic synchronization (phase
synchronization, complete synchronization, lag synchronization)
noise appears to obstruct the synchronous motion and increase the
value of the coupling strength between oscillators corresponding to
the onset of synchronization.

At the same time, effect of noise on the generalized synchronization
regime is investigated poorly enough. As an exception one can refer
to the paper~\cite{Guan:NoiseGS:2006} where the effect of noise on
generalized synchronization in two characteristically different
chaotic oscillators have been considered. In this case the effect of
noise can be system dependent, i.e. common noise can either
induce/enhance or destroy the generalized synchronization regime.

Systems studied in~\cite{Guan:NoiseGS:2006} are close to an
attractor crisis bifurcation~\cite{Grebogi:1982_crisis}. In this
case external noise of small intensity may result in creation of a
new chaotic attractor with a qualitatively different topology that
results in changing of the system behavior in the presence of noise.
In present paper we dwell for the first time upon the behavior of
the generalized synchronization regime in systems which attractors
are far away from the boundary bifurcation crisis or their basins of
attraction are \col{infinitely large}. We report for the first time
theoretical and experimental results of the influence of noise on
the threshold of the generalized synchronization regime in identical
systems with mismatched parameters whose attractors satisfy the
conditions mentioned above. As it would be shown bellow, in this
case the generalized synchronization onset is almost independent on
the noise intensity, i.e. the synchronous regime appears in the
absence and presence of noise practically for the same values of the
coupling parameter strength. At the same time, if the system
attractors are far away from the boundary bifurcation crisis, with
their basins of attraction being limited, the stability of the
generalized synchronization regime \col{with respect} to external
noise would be observed in the large, but limited range of the noise
intensity. The same findings also remain to be correct for the
systems with the \col{infinitely large} basin of attraction,
although the causes of such type of behavior are different.

Revealed peculiarity of the behavior of the boundary of the GS
regime in the presence of noise could be used in many relevant
circumstances, e.g. for the secure transmission of information
through the communication
channel~\cite{alkor:2010_SecureCommunicationUFNeng,Moskalenko:InfoTransNoisePLA2010},
in the medical,
physiological~\cite{Hramov:2006_Prosachivanie,Hramov:2007_UnivariateDataPRE}
and other practical applications where the level of natural noise is
sufficient.

\col{The structure of the paper is as follows}. Section~\ref{sct:GS}
contains brief description of the generalized synchronization
regime, its methods for detection and mechanisms of its arising both
in the cases of the absence and presence of noise. The reasons of
the stability of the generalized synchronization regime \col{with
respect} to noise are also discussed in this section.
Section~\ref{sct:NumExamples} presents results of numerical
simulation of the influence of noise on the threshold of the
synchronous regime arising in several systems with discrete and
continuous time as well as spatially extended systems demonstrating
spatio-temporal chaos. In Section~\ref{sct:Experiment} we describe
the experimental setup for the observation of the generalized
synchronization regime in the presence of noise in the electronic
chaotic circuit and give the results obtained by means of it. Final
discussions and remarks are given in Conclusions.

\section{Generalized synchronization regime}
\label{sct:GS} The generalized synchronization regime (GS) in two
unidirectionally coupled chaotic oscillators \col{with continuous}
\begin{equation}
\begin{split}
&\mathbf{\dot x}(t)=\mathbf{G}(\mathbf{x}(t),\mathbf{g}_d)\\
&\mathbf{\dot u}(t)=\mathbf{H}(\mathbf{u}(t),\mathbf{g}_r)+
\varepsilon\mathbf{P}(\mathbf{x}(t),\mathbf{u}(t)),\\
\end{split}
\label{eq:Oscillators}
\end{equation}
\col{or discrete
\begin{equation}
\begin{split}
&\mathbf{x}_{n+1}=\mathbf{G}(\mathbf{x}_n,\mathbf{g}_d)\\
&\mathbf{u}_{n+1}=\mathbf{H}(\mathbf{u}_n,\mathbf{g}_r)+
\varepsilon\mathbf{P}(\mathbf{x}_n,\mathbf{u}_n),\\
\end{split}
\label{eq:Oscillators_discr}
\end{equation}
time} means the presence of a functional relation
\col{\begin{equation} {\mathbf{u}=\mathbf{F}[\mathbf{x}]}
\label{eq:FunctRel}
\end{equation}}
between the drive $\mathbf{x}$ \col{($\mathbf{x}(t)$ or $\mathbf{x}_n$)} and
response $\mathbf{u}$ \col{($\mathbf{u}(t)$ or $\mathbf{u}_n$)}
system states~\cite{Rulkov:1995_GeneralSynchro,%
Pyragas:1996_WeakAndStrongSynchro}\col{, i.e. in the GS regime the
response system behavior converges to the synchronized state independently on
the choice of initial conditions belonging to the same basin of
attraction}. In equations
(\ref{eq:Oscillators})\col{--(\ref{eq:Oscillators_discr})}
$\mathbf{x}$ and $\mathbf{u}$ are the state vectors of the drive and
response systems, respectively; $\mathbf{G}$ and $\mathbf{H}$ define
the vector fields of interacting systems, $\mathbf{g}_d$ and
$\mathbf{g}_r$ are the control parameter vectors, $\mathbf{P}$
denotes the coupling term, and $\varepsilon$ is the scalar coupling
parameter. \col{Typically, the analytical form of the relation $\mathbf{F}[\cdot]$ in (\ref{eq:FunctRel})
can not be found in most cases.} Depending on the
character of this relation -- smooth or fractal -- GS can be divided
into the strong and the weak
ones~\cite{Pyragas:1996_WeakAndStrongSynchro}, respectively. It is
also important to note that the distinct dynamical systems
(including the systems with the different dimension of the phase
space) may be used as the drive and response oscillators to achieve
the GS regime.

To detect the GS regime \col{both in flow systems and discrete maps}
the auxiliary system method~\cite{Rulkov:1996_AuxiliarySystem} is
frequently used. According to this method the behavior of the
response system $\mathbf{u}$ is considered
together with the auxiliary system $\mathbf{v}$
\col{($\mathbf{v}(t)$ in the case of the flow systems and $\mathbf{v}_n$ if maps are considered)}. The auxiliary system is equivalent to
the response one by the control parameter values, but starts with
other initial conditions belonging to the same basin of chaotic
attractor (if there is the multistability in the system). If GS
takes place, the system states $\mathbf{u}$
and $\mathbf{v}$
become identical after the transient is finished due to the
existence of the relations \col{$\mathbf{u}=\mathbf{F}[\mathbf{x}]$}
and \col{$\mathbf{v}=\mathbf{F}[\mathbf{x}]$}. Thus, the coincidence
of the state vectors of the response and auxiliary systems
\col{$\mathbf{v}\equiv\mathbf{u}$} is considered as a criterion of
the GS regime presence.

It is also possible to compute the conditional Lyapunov exponents to
detect the presence of GS~\cite{Pyragas:1996_WeakAndStrongSynchro}.
In this case Lyapunov exponents are calculated for the response
system, and since the behavior of this system depends on the drive
system these Lyapunov exponents are called conditional. Negativity
of the largest conditional Lyapunov exponent is a criterion of the
GS presence in unidirectionally coupled dynamical
systems~\cite{Pyragas:1996_WeakAndStrongSynchro}.

Methods for the GS regime detection described above could be easily
applied for the investigation of the influence of noise on the GS
regime onset, with all criteria of the GS regime appearance
remaining unchangeable. In other words, the auxiliary system method
and the conditional Lyapunov exponent calculation may be used to
detect the existence of this type of synchronization {both in flow
systems and discrete maps} in the presence of noise.

\col{At the same time, taking into account the fact that the
definition of the GS regime and methods for its detection are the
same both for systems with continuous and discrete time\footnote{Moreover,
flow systems may be reduced to discrete maps, with all types of the
synchronous behavior being connected with each
other~\cite{Filatova:MapsFlowsJETP2005}}, further in this Section we
consider the GS regime onset in
flow systems. Several peculiarities connected with the GS regime
onset in discrete maps will be discussed in
Section~\ref{sct:LogMaps}.}

GS is known to take place in systems with the different types of coupling, the dissipative and
non-dissipative ones. In the case of dissipatively coupled identical
\col{flow} dynamical systems with mismatched parameters
equations~(\ref{eq:Oscillators}) can be rewritten as
\begin{equation}
\begin{split}
&\mathbf{\dot x}(t)=\mathbf{H}(\mathbf{x}(t),\mathbf{g}_d)\\
&\mathbf{\dot u}(t)=\mathbf{H}(\mathbf{u}(t),\mathbf{g}_r)+
\varepsilon\mathbf{A}(\mathbf{x}(t)-\mathbf{u}(t)),
\end{split}
\label{eq:Oscillators1}
\end{equation}
where $\mathbf{A}={\{\delta_{ij}\}}$ is the coupling matrix,
$\delta_{ii}=0$ or $\delta_{ii}=1$, $\delta_{ij}=0$ ($i\neq j$). The
mechanisms of the GS regime arising in systems with the dissipative
coupling can be revealed by the modified system approach firstly
proposed in our previous work~\cite{Aeh:2005_GS:ModifiedSystem}. Due
to such approach the dynamics of the response system may be
considered as the non-autonomous dynamics of the modified system
\begin{equation}
\mathbf{\dot
u}_m(t)=\mathbf{H}'(\mathbf{u}_m(t),\mathbf{g}_r,\varepsilon)
\label{eq:RsOsc}
\end{equation}
where $\mathbf{H}'(\mathbf{u}(t))=
\mathbf{H}(\mathbf{u}(t))-\varepsilon\mathbf{A}\mathbf{u}(t)$, under
the external force $\varepsilon\mathbf{A}\mathbf{x}(t)$:
\begin{equation}
\mathbf{\dot
u}_m(t)=\mathbf{H}'(\mathbf{u}_m(t),\mathbf{g}_r,\varepsilon)+
\varepsilon\mathbf{A}\mathbf{x}(t). \label{eq:RsOsc&Force}
\end{equation}
One can easily see that the term
$-\varepsilon\mathbf{A}\mathbf{x}(t)$ brings the additional
dissipation into the system~(\ref{eq:RsOsc}). The phase flow
contraction is characterized by means of the vector field
divergence. Obviously, the vector field divergences of the modified
and the response systems are related with each other as
\begin{equation}
\mathrm{div}\,\mathbf{H}'=\mathrm{div}\,\mathbf{H}
-\varepsilon\sum\limits_{i=1}^N\delta_{ii}
\end{equation}
(where $N$ is the dimension of the modified system phase space),
respectively. So, the dissipation in the modified system is greater
than in the response one and it increases with the growth of the
coupling strength $\varepsilon$.

The GS regime arising in~(\ref{eq:Oscillators1}) may be considered
as a result of two cooperative processes taking place
simultaneously. The first of them is the growth of the dissipation
in the system~(\ref{eq:RsOsc}) and the second one is an increase of
the amplitude of the external signal. Both processes are correlated
with each other by means of parameter $\varepsilon$ and can not be
realized in the coupled oscillator system~(\ref{eq:Oscillators1})
independently. Nevertheless, it is clear, that an increase of the
parameter $\varepsilon$ in the modified system~(\ref{eq:RsOsc})
results in the simplification of its behavior and the transition
from the chaotic oscillations to the periodic
ones~\cite{Aeh:2005_GS:ModifiedSystem}. Moreover, if the additional
dissipation is large enough the stable fixed state may be realized
in the modified system. On the contrary, the external chaotic force
$\varepsilon\mathbf{A}\mathbf{x}(t)$ tends to complicate the
behavior of the modified system and impose its own dynamics on it.
The GS regime is known to take place when own chaotic dynamics of
the autonomous modified system is
suppressed~\cite{Aeh:2005_GS:ModifiedSystem}. At the same time, the
response system demonstrates chaotic oscillations due to the
external signal \col{coming} from the drive system.

So, the stability of the GS regime is defined primarily by the
properties of the modified system. \col{Adding of noise} does not
change the characteristics of the modified system (\ref{eq:RsOsc})
and does not seem to \col{affect the threshold} of the GS regime
onset. Therefore, the GS regime should exhibit the stability
\col{with respect} to noise in the wide range of the noise
intensities. \col{At that, it should be noted that the
characteristics of noise does not matter and the similar stability
of the GS regime would be observed both for additive and
multiplicative noise with different characteristics.}

To verify the correctness of the statement mentioned above we use
the conditional Lyapunov exponent method. We consider the evolution
of both the reference state of the response oscillator
$\mathbf{u}(t)$ and the perturbed one
${\mathbf{v}(t)=\mathbf{u}(t)+\mathbf{\Delta}(t)}$ being close to
each other (i.e., $|\mathbf{\Delta}(t)|\ll 1$). The conditional
Lyapunov exponents $\lambda_i^r$ ($i=1,\dots,N$) are determined by
the exponential increase/decrease of the small perturbation
$\mathbf{\Delta}(t)$. To take into account the noise influence we
have added the noise terms $\zeta, \xi\in\mathbb{R}^N$ into
equations~(\ref{eq:Oscillators1}) describing the dynamics of the
drive and response systems:
\begin{equation}
\begin{split}
&\mathbf{\dot x}(t)=\mathbf{H}(\mathbf{x}(t),\mathbf{g}_d)+\mathbf{\zeta}(t)\\
&\mathbf{\dot u}(t)=\mathbf{H}(\mathbf{u}(t),\mathbf{g}_r)+
\varepsilon\mathbf{A}(\mathbf{x}(t)-\mathbf{u}(t))+\xi(t).
\end{split}
\label{eq:OscillatorsNoise}
\end{equation}
In Eq.~(\ref{eq:OscillatorsNoise}) the stochastic processes are
supposed to be different for the more complicated case to be
considered.

\col{In this case the dynamics of the auxiliary (perturbed) system
would be given by
\begin{equation}
\mathbf{\dot v}(t)=\mathbf{H}(\mathbf{v}(t),\mathbf{g}_r)+
\varepsilon\mathbf{A}(\mathbf{x}(t)-\mathbf{v}(t))+\xi(t).
\label{eq:PerturbNoise}
\end{equation}
Note, the concept of the generalized synchronization and the auxiliary system approach requires the identity of the signals driving both the response and auxiliary systems. This requirement means that noise must be also identical for the response and auxiliary system. In other words, the control parameter values and the
noise signals in the response and  auxiliary systems should be fully
identical whereas initial conditions for them should be chosen
different.}

\col{The equation determining the evolution of the perturbation
$\mathbf{\Delta}(t)$ may be obtained as follows
\begin{equation}
\begin{array}{lll}
\dot{\mathbf{
\Delta}}(t)=\mathbf{H}(\mathbf{v}(t),\mathbf{g}_r)-\mathbf{H}(\mathbf{u}(t),\mathbf{g}_r)-\varepsilon
A\mathbf{\Delta}(t).
\end{array}
\label{eq:Variation1}
\end{equation}}
Taking into account that
$\mathbf{v}(t)=\mathbf{u}(t)+\mathbf{\Delta}(t)$ and
${|\mathbf{\Delta}(t)|\ll 1}$,  one can write
\begin{equation}
\mathbf{H}(\mathbf{v}(t),\mathbf{g}_r)\approx\mathbf{H}(\mathbf{u}(t),\mathbf{g}_r)+\mathbf{J}\mathbf{H}(\mathbf{u}(t),\mathbf{g}_r)\mathbf{\Delta}(t)
\end{equation}
(where $\mathbf{J}$ is a Jakobian matrix), and, as a consequence
\begin{equation}
\dot{
\mathbf{\Delta}}(t)=(\mathbf{J}\mathbf{H}(\mathbf{u}(t))-\varepsilon\mathbf{A})
\Delta (t)=\mathbf{J}\mathbf{H}'(\mathbf{u}(t))\mathbf{\Delta}(t),
\label{eq:Variation}
\end{equation}
Eq.~(\ref{eq:Variation}) is the variational equation for the
computation of the conditional Lyapunov exponents of the response
system describing by Eq.~(\ref{eq:OscillatorsNoise}) as well as
Eq.~(\ref{eq:Oscillators1}). Therefore, one can conclude that the
largest conditional Lyapunov exponents (determining the threshold of
the GS regime onset) would behave in the similar way both in the
absence and presence of noise. Therefore, the threshold of GS should
not considerably depend on the noise intensity, whereas the GS
regime should exhibit the stability to the noise influence. Note,
also, that the vector state ${\mathbf{u}(t)}$ in
Eq.~(\ref{eq:Variation}) depends on the noise signals, and,
therefore, \col{the largest conditional Lyapunov exponents obtained for the cases with and without noise are, however, not equivalent. As a consequence, } the great intensities of noise may change the stability
properties of the modified system that may result in the variation
of the value of the threshold of the GS regime.

It should be noted that onset of the GS regime is similar to the
last one for the cases of complete (CS) (identical) and lag (LS)
synchronization. Such types of the synchronous chaotic system
behavior could be considered as partial cases of GS and they
correspond to the stronger forms of such
regime~\cite{Pyragas:1996_WeakAndStrongSynchro}. It is clear that
the modified system approach could be applied for revealing the
mechanisms resulting in the synchronous regime onset even in the
presence of noise. At the same time, it should be noted that even
for identical dynamical systems with equal values of the control
parameters GS regime arises a bit earlier than the CS
one~\cite{Pyragas:1996_WeakAndStrongSynchro,Harmov:2005_GSOnset_EPL}.
As it would be shown bellow in Section~\ref{sct:Roessler}, external
noise added to the drive and response could destroy the CS (or LS)
regime but it does not destruct the GS regime itself. Therefore, the
stability of the CS and LS regimes is \col{less strong} than for the
GS one.

\section{Influence of noise on the GS regime onset in sample chaotic systems: numerical calculations}
\label{sct:NumExamples} To illustrate the stability of the GS regime
\col{with respect} to noise we consider numerically three different
pairs of unidirectionally dissipatively coupled chaotic dynamical
systems being capable to demonstrate the GS regime. As such model
systems we have selected (i) systems with discrete time -- two
unidirectionally coupled logistic maps, (ii) chaotic oscillators --
two unidirectionally coupled R\"ossler systems; (iii) spatially
extended dynamical systems -- unidirectionally coupled
one-dimensional complex Ginzburg-Landau equations.

\subsection{Logistic maps}
\label{sct:LogMaps} We start our consideration with the GS regime
arising in two unidirectionally coupled logistic maps with
\col{additive} noise term:
\begin{equation}
\begin{array}{l}
x_{n+1}=f(x_n,\lambda_x),\\
y_{n+1}=f(y_n,\lambda_y)+\varepsilon(f(x_n,\lambda_x)+Df(\xi_n,\lambda_x)-f(y_n,\lambda_y)),
\end{array}
\label{eq:LogMapNoise}
\end{equation}
where $f(x,\lambda)=\lambda x(1-x)$, ${\varepsilon<1}$. Here $\lambda_{x,y}$ are the
control parameter values of the drive and response systems,
respectively, $\varepsilon$ characterizes the coupling strength
between systems, $\xi_n$ is the stochastic process which probability
density is distributed uniformly on the interval $[0;1]$, $D$
defines the intensity of added noise.

\col{Although the systems with the discrete time are the specific class of dynamical systems, they are closely interrelated with the flow systems~\cite{Starodubov:2006_MapsAndFlows}, with types of the chaotic synchronous motion corresponding with each other in maps and flows~\cite{Filatova:MapsFlowsJETP2005}. Nevertheless, there are also differences between these types of chaotic dynamical systems. One of them is the type of coupling between oscillators. Typically, for the logistic maps the coupling term is introduced in the same way as it has been done in Eq.~(\ref{eq:LogMapNoise}) instead of the linear difference of the vectors (like in Eq.~(\ref{eq:Oscillators1})), since for the maps it is this kind of terms that provides the dissipative type of coupling~\cite{Pyragas:1996_WeakAndStrongSynchro,Aeh:2005_GS:ModifiedSystem,Hramov:2006_PLA_NIS_GS}). Additionally, here and later the noise signal is introduced in the
coupling term to emulate a natural noise added in the communication
channel~\cite{Moskalenko:InfoTransNoisePLA2010}}.
To detect the GS regime in such system we have computed
conditional Lyapunov exponents with further refinement of the
threshold values by the auxiliary system method described above.

\begin{figure}[tb]
\centerline{\scalebox{0.45}{\includegraphics{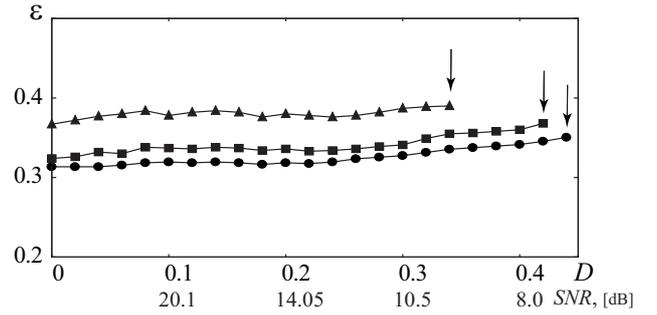}}}
\caption{Dependence of the threshold of the GS regime onset in two
unidirectionally coupled logistic maps (\ref{eq:LogMapNoise}) on the
intensity of noise (the SNR values corresponding to the noise
intensities are also shown) for different values of the control
parameters: $\lambda_x=3.75$, $\lambda_y=3.75$ ({\Large $\bullet$}),
$\lambda_x=3.75$, $\lambda_y=3.79$ ($\blacksquare$),
$\lambda_x=3.75$, $\lambda_y=3.9$ ({\large$\blacktriangle$}).
Critical values of the noise intensity $D_c$, up to which the GS
regime in system (\ref{eq:LogMapNoise}) is observed, are marked by
arrows \label{fgr:LogMapNoise}}
\end{figure}

The dependence of the GS regime onset on the noise intensity $D$ for
different values of the control parameters $\lambda_{x,y}$ is shown
in Fig.~\ref{fgr:LogMapNoise}.
On the horizontal axis the signal to noise ratios (SNR, [dB])
corresponding to these noise intensities are also
indicated~\footnote{Here and later in the paper the SNR value has
been computed in traditional way, i.e. $\displaystyle
\rm{SNR}=10\lg\frac{P_{sign}}{P_{noise}}$, where $\rm P_{sign}$ is a
power of chaotic signal, $\rm P_{noise}$ is a power of noise
\col{affected the chaotic system}~\cite{Sklyar:CifrCommun2003Eng}.
\col{The power of the signal $x(t)$ (independently of the fact
whether it is deterministic or stochastic) on the time interval
$[0;T]$ has been computed by its time realization, i.e.
$P=\int_0^Tx^2(t)dt$.}}. One can easily see that the threshold value
of the coupling parameter $\varepsilon$ is almost independent on the
intensity of noise $D\in[0;D_c]$ where $D_c$ shown in
Fig.~\ref{fgr:LogMapNoise} by arrows, depends on the control
parameter values, $D_c=D_c(\lambda_x,\lambda_y)$. For the selected
values of the control parameters $D_c\in[0.38;0.44]$ ($\rm
SNR\in[8.5;7.2]dB$, respectively), i.e. the GS regime in
unidirectionally coupled logistic maps (\ref{eq:LogMapNoise}) is
stable to noise up to the power of noise comparable with the chaotic
signal one.

\begin{figure}[b]
\centerline{\scalebox{0.45}{\includegraphics{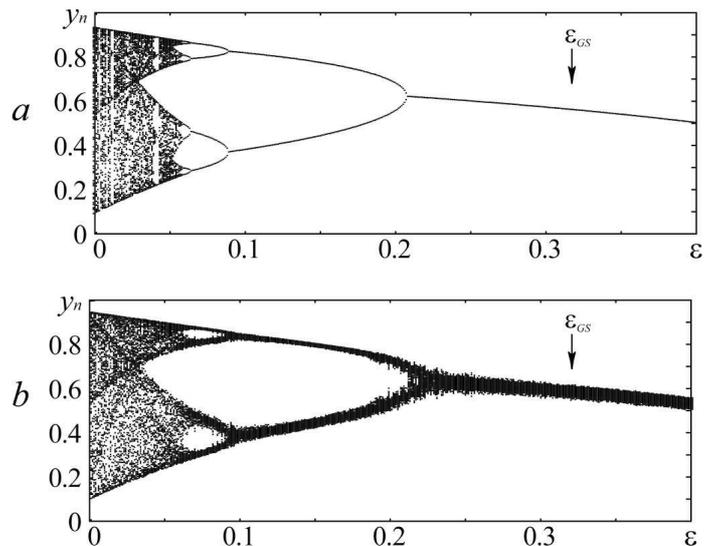}}}
\caption{Bifurcation diagrams of the modified logistic
map~(\ref{eq:ModLogMap}) in the absence (\textit{a}) and presence
(\textit{b}) of noise \col{(the noise is introduced in system (\ref{eq:ModLogMap}) in the
same way as in Eq.~(\ref{eq:LogMapNoise}))}. The control parameter $\lambda_y=3.79$ in
both cases, the noise intensity $D=0.1$ in (\textit{b}). The
coupling parameter values $\varepsilon_{GS}=0.32$ corresponding to
the GS regime \col{(obtained by means of conditional Lyapunov exponent computation)} are marked by arrow in both cases
\label{fgr:BifLogMod}}
\end{figure}

To explain the reasons of stability of the GS regime \col{with
respect} to external noise we \col{use the modified system approach
described in Section~\ref{sct:GS}. At the same time, due to the fact
that the theory of the stability of the GS regime to noise proposed
in Section~\ref{sct:GS}, is applicable to flow systems, and the
noise added in system (\ref{eq:LogMapNoise}) is multiplicative,
there are several peculiarities to be discussed bellow. Therefore,
we use the modified system approach with regard to the system with discrete time and} consider the modified logistic map:
\begin{equation}
z_{n+1}=\col{f_m(z_n,\lambda_y)=}(1-\varepsilon)f(z_n,\lambda_y).
\label{eq:ModLogMapGeneral}
\end{equation}
\col{One can see that the modified system~(\ref{eq:ModLogMapGeneral}) may be rewritten in the form
\begin{equation}
z_{n+1}=az_n(1-z_n),
\label{eq:ModLogMap}
\end{equation}
}
where $a=\lambda_y(1-\varepsilon)$. \col{It is clearly seen that the
term $-\varepsilon f(z_n,\lambda_y)$ brings additional dissipation
in system~\col{(\ref{eq:ModLogMapGeneral})}. The local phase volume contraction is
characterized by means of the modulus of the derivative
\begin{equation}
\left|\displaystyle \frac{dz_{n+1}}{dz_n}\right|=(1-\varepsilon)\left|f'_{z_n}(z_n,\lambda_y)\right|
\label{eq:LogMapContr}
\end{equation}
where the modulus of multiplier ${|\displaystyle
f'_{z_n}(z_n,\lambda_y)|}=\lambda_y|1-2z_n|$ characterizes the
phase volume contraction in the autonomous response system. The case of ${\left|dz_{n+1}/dz_n\right|=1}$ corresponds to the non-dissipative dynamics whereas $\left|{dz_{n+1}}/{dz_n}\right|=0$ relates to the infinitely large dissipation. One can see that, as
in the case of flow systems, the dissipation in the modified system
is greater than in the response one and it increases with the growth
of the coupling strength $\varepsilon$, ($0<\varepsilon<1$).} Bifurcation diagram
characterizing its behavior with the increase of
$\varepsilon$-parameter is shown in
Fig.~\ref{fgr:BifLogMod},\textit{a}.
The value of parameter $\varepsilon$ corresponding to the onset of
the GS regime \col{(obtained by means of conditional Lyapunov exponent computation, see Sec.~\ref{sct:GS})} is marked by arrow. One can see that for a coupling
parameter strengths corresponding to the onset of the GS regime in
system~(\ref{eq:LogMapNoise}), in full agreement with the arguments
discussed in~\cite{Aeh:2005_GS:ModifiedSystem}, the modified
system~(\ref{eq:ModLogMap}) demonstrates the periodic oscillations.
External noise does not almost change the characteristics of the
modified system and, therefore, it does not \col{affect the
threshold} of the GS regime arising. Bifurcation diagram of the
modified logistic map in the presence of noise of intensity $D=0.1$
is shown in
Fig.~\ref{fgr:BifLogMod},\textit{b}.
\col{The noise is introduced in system (\ref{eq:ModLogMap}) in the
same way as in Eq.~(\ref{eq:LogMapNoise}), i.e.
\begin{equation}
z_{n+1}=f_m(z_n,\lambda_y)+\varepsilon Df(\xi_n,\lambda_x)
\label{eq:ModifiedLogMapNoise}
\end{equation}
to provide the same noise influence as in the coupled logistic maps.}
The level of noise is quite
sufficient in comparison with the signal amplitude what is clearly
seen from the kind of bifurcation diagram. At the same time, it is
easy to see that noise does not shift the bifurcation points in this
case but only leads to a noisiness of the system regime.
Therefore, in the considered case one can say that, despite of the
large amplitude, the external noise does not \col{affect the
threshold} of the GS regime onset. The further increase of the noise
intensity $D>D_c$ results in the runaway of the representation point
to infinity.

The reasons of the jump of the representation point to infinity can
be explained in the following way. Logistic map in autonomous regime
\begin{equation}
x_{n+1}=f(x_n,\lambda), \label{eq:LogMapAut}
\end{equation}
is known to have a finite basin of attraction, i.e., depending on
the choice of the initial conditions, for the values of the control
parameter $\lambda$ mentioned above it demonstrates either chaotic
regime or \col{the jump of representation point to
infinity}~\cite{Schuster:1984_CHAOS_BOOK}. To provide the chaotic
regime in system (\ref{eq:LogMapAut}) we have to specify initial
condition in range $x_0\in[0;1]$, with the representation point
remaining in this range during the evolution of the system for an
infinitely long time\col{, at that the maximal value of
$f(x_n)=f_{max}=\lambda/4$ would be achieved if $x_n=1/2$.} At the
same time, it is clear that external noise could make it go out the
range mentioned above.

The similar effect takes place for systems~(\ref{eq:LogMapNoise}).
One can estimate the intensity of noise $D_c$ corresponding to the
\col{jump of representation point of the response system to
infinity. For this purpose we consider the behavior of the drive and
response systems of Eq.~(\ref{eq:LogMapNoise}). First equation
corresponds to the drive system and is not affected by the influence
of external noisy or chaotic signal. Therefore the most probable
value of $f(x_n)=\langle f_x\rangle$ where $\langle f_x\rangle$ is a
statistical mean of $f(x_n)$ (due to the properties of the
autonomous logistic map). The maximal values of $f(y_n)$ and
$f(\xi_n)$ would be equal to $f_{max}$ because of the uniform
character of the probability distribution of the random value
$\xi_n$ and the arguments discussed above. Due to the fact that a
random variable $\xi_n$ could not be negative the jump of the
representation point from the range $[0;1]$ could be performed only
through a right boundary of such range. Therefore the maximal value
of $y_{n+1}$ is $1$. Substituting all quantities into the second equation
of (\ref{eq:LogMapNoise}) and assuming $\varepsilon=\varepsilon_c$
($\varepsilon_c$ corresponds to the threshold value of the GS regime
onset without noise) we can estimate the approximate values of the
noise intensity $D_c$ \col{up to which the jump of representation
point to infinity does not take place.}  Our calculations show that}
\begin{equation}
D_c\approx\frac{4-4\varepsilon_c\langle
f_x\rangle-(1-\varepsilon_c)\lambda_y}{\varepsilon_c\lambda_x},
\label{eq:Dc1}
\end{equation}
i.e. ${D_c\approx0.5}$ for the control parameter values
${\lambda_x=\lambda_y=3.75}$, ${D_c\approx0.48}$ for
${\lambda_x=3.75}$, ${\lambda_y=3.79}$ and ${D_c\approx0.42}$ for
${\lambda_x=3.75}$, ${\lambda_y=3.9}$, that agrees well with the
results of direct numerical calculations. Therefore, the GS regime
for logistic maps, having a limited basin of attraction, exhibit the
stability \col{with respect} to noise in the large, but limited
range of the noise intensity.

In the considered case the GS regime destruction is connected with
the \col{jump of representation point to infinity}, which could be
considered as an attraction of it to the second coexisting attractor
being at the infinity~\cite{Grebogi:1983_Basins}. Note, if the
coexisting attractor was characterized by the limited basin of
attraction, depending on the type of the regime being realized in
the response system (and, correspondingly, to the second attractor),
the increase or decrease of the threshold value of the GS regime
would be observed with the growth of the noise
intensity~\cite{Guan:NoiseGS:2006}.

The another important question is the stability of GS \col{with
respect} to the external noise in the case when statistically
independent noise sources \col{affect the drive and response
systems}
\begin{equation}
\begin{array}{l}
x_{n+1}=f(x_n,\lambda_x)+\varepsilon Df(\zeta_n,\lambda_x),\\
y_{n+1}=f(y_n,\lambda_y)+\varepsilon(f(x_n,\lambda_x)+Df(\xi_n,\lambda_x)-f(y_n,\lambda_y)),
\end{array}
\label{eq:LogMapNoise2}
\end{equation}
where $\zeta_n$ is a stochastic process with the probability density
distributed uniformly in $[0;1]$-range. \col{Applying the arguments
similar to the last one described for the system
(\ref{eq:LogMapNoise}) to system (\ref{eq:LogMapNoise2}), we can
estimate the intensity of noise $D_c^d$ corresponding to the
\col{jump of the representation point of the drive system to
infinity}. It is clear that due to the absence of the dissipative
term in the drive system \col{the jump of representation point  in
it would take place for a less values of the noise intensity than
for the response one.}} In this case the GS regime is stable to the
noise influence until ${D<D_c^d}$, where
\begin{equation}
\displaystyle
D_c^d\approx\frac{1-\lambda_x/4}{\varepsilon_c\lambda_x/4}\approx 0.2.
\label{eq:Dc2}
\end{equation}
The further increase of the noise intensity $D>D_c^d$ results in the
chaotic regime destruction in the drive system. Therefore, the
values $D>D_c^d$ are unapplicable for (\ref{eq:LogMapNoise2}), since
the jump of the representation point towards infinity in the drive
system is observed.

\begin{figure}[tb]
\centerline{\scalebox{0.45}{\includegraphics{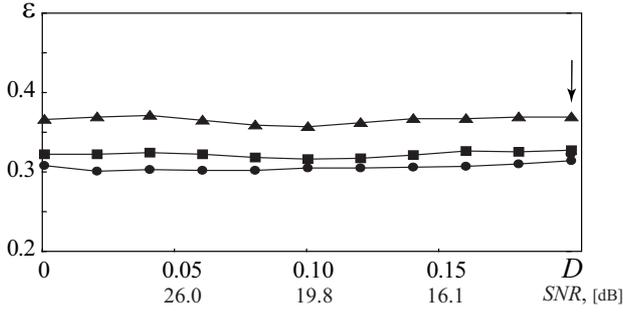}}}
\caption{Dependence of the threshold of the GS regime onset in two
unidirectionally coupled logistic maps (\ref{eq:LogMapNoise2}) in
the presence of noise both in the drive and response on its
intensity (the SNR values corresponding to the noise intensities are
also shown) for different values of the control parameters:
$\lambda_x=3.75$, $\lambda_y=3.75$ ({\Large $\bullet$}),
$\lambda_x=3.75$, $\lambda_y=3.79$ ($\blacksquare$),
$\lambda_x=3.75$, $\lambda_y=3.9$ ({\large$\blacktriangle$})
\label{fgr:LogMapNoise2}}
\end{figure}

Numerical calculations confirm the results obtained analytically. In
Fig.~\ref{fgr:LogMapNoise2} dependencies of the critical values of
the coupling parameter strength corresponding to the GS regime
arising for different values of the control parameters are shown
(the SNR values are also indicated in the second horizontal axis).
As in the case of the absence of noise in the drive system external
noise does not \col{affect the threshold} value of the GS regime
onset. The very similar result is obtained in the case when both the
drive and response systems are under the influence of the common
noise source, i.e., $\xi_n\equiv\zeta_n$.

It should be noted that the GS regime is also robust in the limited
range against the perturbations in the control parameters by noise.
Therefore, one can conclude that for unidirectionally dissipatively
coupled systems with discrete time the GS regime would exhibit the
stability \col{with respect} to noise.

\subsection{R\"ossler systems} \label{sct:Roessler}
As a second example we consider two unidirectionally coupled flow
R\"ossler oscillators:
\begin{equation}
\begin{array}{ll}
\dot x_1=-\omega_{x}x_2-x_3+D_1\zeta,\\
\dot x_2=\omega_{x}x_1+ax_2,\\
\dot x_3=p+x_3(x_1-c),\\
\\
\dot u_1=-\omega_{u}u_2-u_3 +\varepsilon(x_1+D_2\xi-u_1),\\
\dot u_2=\omega_{u}u_1+au_2,\\
\dot u_3=p+u_3(u_1-c),
\end{array}
\label{eq:Roesslers}
\end{equation}
where $\mathbf{x}(t)=(x_1,x_2,x_3)^T$ and
$\mathbf{u}(t)=(u_1,u_2,u_3)^T$ are the vector-states of the drive
and response systems, respectively, $a=0.15$, $p=0.2$, $c=10$,
$\omega_x$ and $\omega_u=0.95$ are the control parameter values,
$\varepsilon$ is a coupling parameter. The parameters $\omega_{x,u}$
define the natural frequencies of the drive and response system
oscillations. The terms $D_1\zeta$, $D_2\xi$ simulate the external
noise influenced on the drive and response systems. Here $\xi$ and
$\zeta$ are statistically independent stochastic Gaussian processes
described by the following probability distribution
\begin{equation}
p(\xi)=\frac{1}{\sqrt{2\pi}\sigma}\exp\left(-\frac{(\xi-\xi_0)^2}{2\sigma^2}\right),
\label{eq:NormalDistrib}
\end{equation}
where $\xi_0=0$ and $\sigma=1.0$ are the mean value and variance.
Parameters $D_{1,2}$ define the intensities of the noise added in
the drive and response systems, respectively.

To integrate the stochastic equations~(\ref{eq:Roesslers}) we have
used the four order Runge-Kutta method adapted for the stochastic
differential equations~\cite{Nikitin:1975_StochDE_eng} with time
discretization step $\Delta t=0.001$. The modified Runge-Kutta
method is applicable for delta-correlated Gaussian white noise used
frequently in our Manuscript. At the same time, for the integration
of the stochastic differential equations in the case of the other
types of noise we have used one-step Euler method. For the GS regime
detection the auxiliary system method described in
Section~\ref{sct:GS} has been used.

\begin{figure}[tb]
\centerline{\scalebox{0.28}{\includegraphics{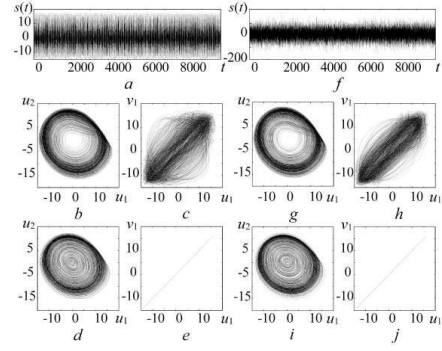}}}
\caption{Signals $s(t)$ \col{affecting the response} and auxiliary
systems (\textit{a,f}), phase portraits (\textit{b,d,g,i}) and
$(u_1,v_1)$-planes characterizing the response and auxiliary system
behavior (\textit{c,e,h,j}) before ($\varepsilon=0.05$) and after
($\varepsilon=0.114$) the GS regime onset in unidirectionally
coupled R\"ossler systems with $\omega_d=0.99$, respectively.
Pictures (\textit{a--e}) correspond to the noiseless case ($D=0$)
whereas (\textit{f--j}) refer to the noise one ($D=40$)
\label{fgr:PhasePortNoiseless}}
\end{figure}

\begin{figure}[tb]
\centerline{\scalebox{0.45}{\includegraphics{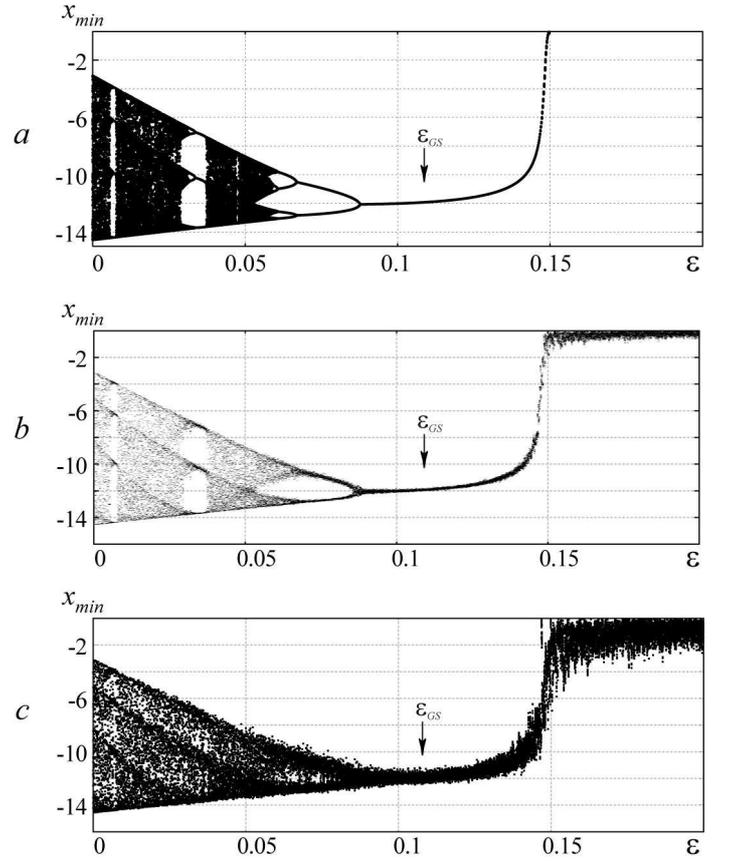}}}
\caption{Bifurcation diagrams of the modified R\"ossler
system~(\ref{eq:ModRoesslers}) in the absence (\textit{a}) and
presence (\textit{b,c}) of noise \col{(the noise is introduced in system (\ref{eq:ModRoesslers}) in the
same way as in Eq.~(\ref{eq:Roesslers}))}. The control parameter
$\omega_x=0.99$ in all considered cases, the noise intensity $D=10$
in (\textit{b}) and $D=40$ in (\textit{c}). The coupling parameter
values $\varepsilon_{GS}=0.112$ corresponding to the GS regime \col{(obtained by means of auxiliary system method, see Sec.~\ref{sct:GS})} are
marked by arrow in all cases \label{fgr:BifRossMod}}
\end{figure}

At first, we consider the behavior of chaotic systems
(\ref{eq:Roesslers}) in the presence of noise influenced only on the
response system, i.e. $D_1=0$, $D_2=D$.
\begin{figure}[t]
\centerline{\scalebox{0.45}{\includegraphics{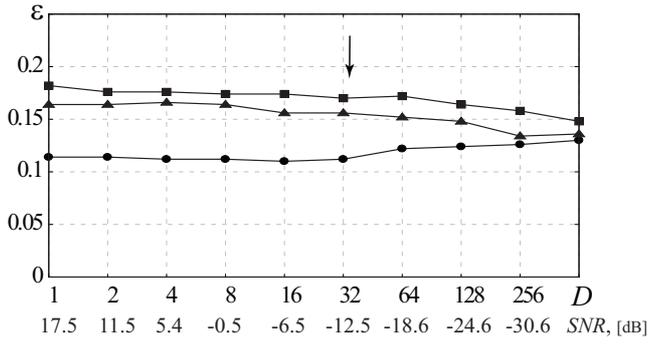}}}
\caption{Dependence of the boundary value corresponding to the GS
regime arising in two unidirectionally coupled R\"ossler systems
with \col{additive} stochastic term~(\ref{eq:Roesslers}) on the noise
intensity $D$ (the SNR values corresponding to the noise intensities
are also shown) for different values of the drive system parameter
$\omega_x$: $\omega_x=0.99$ ({\Large $\bullet$}), $\omega_x=0.95$
($\blacksquare$), $\omega_x=0.91$ ({\large$\blacktriangle$}). The
critical value of the noise intensity $D_c$ up to which the
\col{boundary value of the GS regime does not almost depend on the noise intensity}
is marked by arrow \label{fgr:RoesNoise}}
\end{figure}
Fig.~\ref{fgr:RoesNoise} shows the dependence of the threshold of
the GS regime onset on the noise amplitude $D$ (the SNR value) for
three different values of the control parameter $\omega_x$ and fixed
values of the other control parameters. To possess all necessary
knowledge about influence of noise on the system under study we have
chosen values of the parameter $\omega_x$ in the different ranges of
the parameter mismatch where the different mechanisms of the
synchronous regime arising have been shown to take
place~\cite{Harmov:2005_GSOnset_EPL}. Parameter $\omega_x=0.99$
corresponds to the case of the relatively large values of the
frequency detuning whereas $\omega_x=0.95$ (\col{interacting}
systems are identical) and $\omega_x=0.91$ relate to the small ones.
It is easy to see that independently on the value of the control
parameter $\omega_x$ the threshold of the GS regime onset does not
almost depend on the noise amplitude $D\in[0;40]$ (${\rm
SNR>-14.5dB}$). Even for a great values of the noise intensity GS
arises practically for the same values of the coupling parameter
strength $\varepsilon$ as for a noiseless case. Typical signals
$s(t)=x_1(t)+D\xi(t)$ \col{affecting the response} and auxiliary
systems both in the absence and presence of noise as well as the
phase portraits of the response system and $(u_1,v_1)$-planes
characterizing the response and auxiliary system behavior before
(\textit{b,c,g,h}) and after (\textit{d,e,i,j}) the GS regime onset
are shown in Fig.~\ref{fgr:PhasePortNoiseless}.
Pictures (\textit{a--e}) correspond to the noiseless case whereas
(\textit{f--j}) refer to the presence of noise of great intensity
$D=40$ \col{affecting the response} system (in the last case the
signal is similar to the stochastic one, with its amplitude being in
approximately 10 times more in comparison with the noiseless case,
compare Fig.~\ref{fgr:PhasePortNoiseless},\textit{a,f}). One can
easily see that characteristics of the response systems are changed
slightly with the noise intensity increasing (compare pictures
\textit{b,d} and \textit{g,i}, respectively) and the boundary value
of the GS regime remains practically the same. The causes
determining the stability of the GS regime \col{with respect} to the
external noise influence are the same as in the already considered
case of the logistic maps (\ref{eq:LogMapNoise}) and could also be
explained by the modified system approach. One can say that for
unidirectionally coupled R\"ossler systems the noise of great
intensity does not change the characteristics of the modified system
\begin{equation}
\begin{array}{l}
\dot z_1=-\omega_{u}z_2-z_3 -\varepsilon z_1,\\
\dot z_2=\omega_{u}z_1+az_2,\\
\dot z_3=p+z_3(z_1-c),
\end{array}
\label{eq:ModRoesslers}
\end{equation}
where $\mathbf{z}=(z_1,z_2,z_3)^T$ is the vector state of the
modified system and, as a consequence, of the response one. By the
analogy with the logistic maps the bifurcation diagrams for the
modified R\"ossler system are shown in Fig.~\ref{fgr:BifRossMod}.
Fig.~\ref{fgr:BifRossMod},\textit{a} corresponds to the noiseless
case ($D=0$) whereas in Fig.~\ref{fgr:BifRossMod},\textit{b,c} the
modified R\"ossler systems with \col{additive} noise of the
different intensities ($D=10$ and $D=40$, respectively) \col{(the
noise is introduced in system (\ref{eq:ModRoesslers}) in the same
way as in Eq.~(\ref{eq:Roesslers}))} are shown. Independently on the
noise intensity for the selected values of the control parameters
the cycle-1 periodic oscillations are observed in the modified
system (\ref{eq:ModRoesslers}) (see
also~\cite{Aeh:2005_GS:ModifiedSystem}).

The external noise does not shift the bifurcation points and,
therefore, does not \col{affect the boundary} value of the GS
regime. Therefore, we can conclude that the mechanisms determining
the GS regime stability are the same as for the system with discrete
time (\ref{eq:LogMapNoise}). At the same time, since the basin of
attraction in the R\"ossler system is unbounded, the effect of the
GS regime destruction described above in Section~\ref{sct:LogMaps}
could not be observed.

\begin{figure}[tb]
\centerline{\scalebox{0.4}{\includegraphics{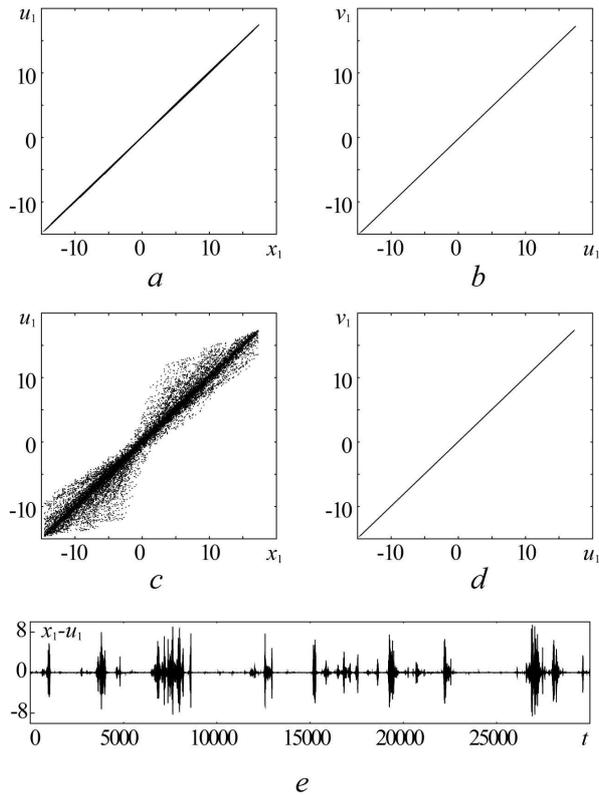}}}
\caption{$(x_1;u_1)$- and $(u_1;v_1)$-planes characterizing the
drive and response (\textit{a,c}) and response and auxiliary
(\textit{b,d}) R\"ossler system behavior in the case of the absence
(\textit{a,b}) and the presence (\textit{c,d}) of noise in the
response system (the noise intensity ${D=0.1}$, the coupling
strength ${\varepsilon=0.19}$). The difference between the drive and
response system states in the presence of noise (${D=0.1}$,
${\varepsilon=0.19}$) shown in Fig.~\ref{fgr:CSon-off},\,\textit{e}
illustrates the presence of on-off intermittency. Parameter of the
drive system ${\omega_d=0.95}$ \label{fgr:CSon-off}}
\end{figure}

One more interesting question to be discussed is the relationship
between the onset of the GS and CS regimes. According to the
consideration made on the base of the modified system approach, GS
and CS have the same mechanisms. At the same time, as we have
mentioned in Section~\ref{sct:GS}, the stability of the GS regime is
stronger than the CS one. To confirm this statement  we have
analyzed the CS regime arising in unidirectionally coupled identical
R\"ossler systems~(\ref{eq:Roesslers}) with $\omega_d=\omega_r=0.95$
and compared obtained results with the last one for the GS. Our
calculations show that in the absence of noise CS arises in this
case for $\varepsilon=0.19$, whereas GS takes place for
$\varepsilon\geq 0.184$. Adding noise of small intensity $D=0.1$
results in the appearance of on-off
intermittency~\cite{Ott:1994_BlowoutBifurcation} between the drive
and response systems, at that the threshold value of the CS regime
grows up, but the GS regime is still observed (see
Fig~\ref{fgr:RoesNoise} and Fig.~\ref{fgr:CSon-off}).

For the very large values of the noise intensity when the power of
noise is much more than the R\"ossler system signal one ($D\gtrsim
400$, $\rm SNR\lesssim -34.5$) the detected synchronous regime may
be treated as the noise-induced synchronization, being the
manifestation of the GS regime in the case when stochastic signal
instead of the deterministic one is \col{affected the response} and
auxiliary systems~\cite{Hramov:2006_PLA_NIS_GS}. In other words, the
deterministic signal from the drive system practically does not play
role and may be neglected in comparison with the stochastic one. At
that, the boundary value of the synchronous regime onset should not
depend on the control parameter of the drive system $\omega_x$ (see,
Fig.~\ref{fgr:RoesNoise} for a large $D$) and is determined mainly
by the characteristics of the noise signal. Therefore, for the noise
intensities $D\gtrsim D_c=45$ ($\rm SNR<-15.5dB$) (shown in
Fig.~\ref{fgr:RoesNoise} by arrow) the threshold value of the
synchronous regime may start increasing or decreasing depending on
the value of the control parameter detuning.

\begin{figure}[tb]
\centerline{\scalebox{0.45}{\includegraphics{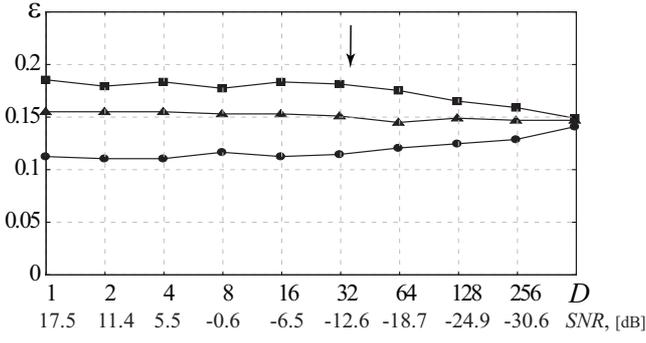}}}
\caption{Dependence of the boundary value of the GS regime arising
in two unidirectionally coupled R\"ossler systems in the case when
statistically independent noise sources of intensity $D$ \col{affect the drive} and response on the noise intensity $D$ (the SNR values
corresponding to the noise intensities are also shown) for different
values of the drive system parameter $\omega_x$: $\omega_x=0.99$
({\Large $\bullet$}), $\omega_x=0.95$ ($\blacksquare$),
$\omega_x=0.91$ ({\large$\blacktriangle$}). The
critical value of the noise intensity $D_c$ up to which the
\col{boundary value of the GS regime does not almost depend on the noise intensity}
is marked by arrow \label{fgr:RoesNoise2}}
\end{figure}

It should be noted that the weak dependence of the threshold value
of the GS regime onset in the wide range of the noise intensity $D$
takes place if the amplitude $D_1$ of the \col{additive} noise term
in equations~(\ref{eq:Roesslers}) is not equal to zero. We have
chosen it to be equal to $D_1=\varepsilon D$. These dependencies for
a different values of the drive system parameter $\omega_x$ are
shown in Fig.~\ref{fgr:RoesNoise2}. Such behavior of
\col{interacting} systems in the presence of noise is fully defined
by mechanisms described above in this subsection.

Therefore, one can say that in both considered cases (maps and
flows) in the wide range of the noise intensity the external noise
does not practically \col{affect the threshold} of the GS regime
arising. Hence, we can say about stability of the GS regime
\col{with respect} to external noise in dynamical systems with a few
number of degrees of freedom.

\subsection{Ginzburg-Landau equations}
\label{sct:GLE} As a third example we consider the GS regime arising
in spatially extended self-oscillating media described by the
complex Ginzburg-Landau equations (CGLE). The system under study is
represented by a pair of unidirectionally dissipatively coupled
complex Ginzburg-Landau equations (CGLE's) being under influence of
distributed in space source of the white noise. Equations describing
such system may be written as
\begin{equation}
\begin{array}{l}
\displaystyle \frac{\partial u}{\partial t}= u-(1-i\alpha_d)|u|^2
u+(1+i\beta_d)\frac{\partial^2 u}{\partial x^2}+ \\ \qquad +
\varepsilon \tilde D\xi(x,t), \quad x\in[0,L],
\end{array}
\label{eq:CGLE_drive}
\end{equation}
\begin{equation}
\begin{array}{l}
\displaystyle \frac{\partial v}{\partial t}= v-(1-i\alpha_r)|v|^2
v+(1+i\beta_r)\frac{\partial^2 v}{\partial x^2} + \\ \qquad +
\varepsilon(\tilde D\zeta(x,t)+u-v), \quad x\in[0,L].
\end{array}
\label{eq:CGLE_response}
\end{equation}
Equation (\ref{eq:CGLE_drive}) describes the drive system and
equation (\ref{eq:CGLE_response}) corresponds to the response one.
It is known that in two unidirectional CGLE's the GS regime may take
place~\cite{Hramov:2005_GLEsPRE}. In our investigation the
parameters of the drive system are chosen as $\alpha_d=1.5$,
$\beta_d=1.5$. To study the generalized synchronization of the
nonidentical systems we have chosen the different values of control
parameters $\alpha_r\in[3;5]$ and $\beta_r\in[3;5]$ for the response
system~(\ref{eq:CGLE_response}). The choice of such values of the
control parameters results in the autonomous systems being in the
spatiotemporal chaotic regime. Parameter $\varepsilon$ determines
the strength of the unidirectionally dissipative coupling between
the response and drive systems, with the interaction of them being
in each point of space. The terms $\tilde D\xi(x,t),\zeta(x,t)$
simulate complex model noise with Gaussian distribution of the
random values $\xi(x,t),\zeta(x,t)$ with zero mean value:
\begin{equation}
\begin{array}{ll}
\langle\zeta(x,t)\rangle=0, \\
\langle\zeta(x,t)\zeta(x',t')\rangle=\delta(x-x')\delta(t-t'),
\end{array}
\label{eq:noise:01}
\end{equation}
$\tilde D$ defines the noise intensity.

Equations~(\ref{eq:CGLE_drive})--(\ref{eq:CGLE_response}) have been
solved with periodic boundary conditions ${u(x,t)=u(x+L,t)}$ and
${v(x,t)=v(x+L,t)}$, with all numerical calculations being performed
for a fixed system length $L=40\pi$ and random initial conditions.
To evaluate (\ref{eq:CGLE_drive})--(\ref{eq:CGLE_response}) the
standard numerical scheme for integration of the stochastic partial
differential equations~\cite{Garcia-Ojalvo:1999_NoiseBook} has been
used, the value of the grid spacing is $\Delta x=L/1024$, the time
step of the scheme is $\Delta t=0.0002$.

To detect the presence of the GS regime we have used the auxiliary
system method described in Section~\ref{sct:GS}. At that, we have
assumed that auxiliary system $v_a(t)$, also
satisfying~(\ref{eq:CGLE_response}), has been under influence of the
noise source of the intensity $\tilde D$ equal to the last one for
the response system. As an criterion of the GS regime arising we
have chosen the following one. The GS regime takes place when the
mean standard deviation of the response $v$ and auxiliary $v_a$
system states satisfies the following condition:
\begin{equation}
\frac{1}{T}\int_T\int_0^L |v(x,t)-v_a(x,t)|^2<\delta,
\label{eq:noise:02}
\end{equation}
where $\delta=10^{-5}$.

\begin{figure}[tb]
\centerline{\scalebox{0.45}{\includegraphics{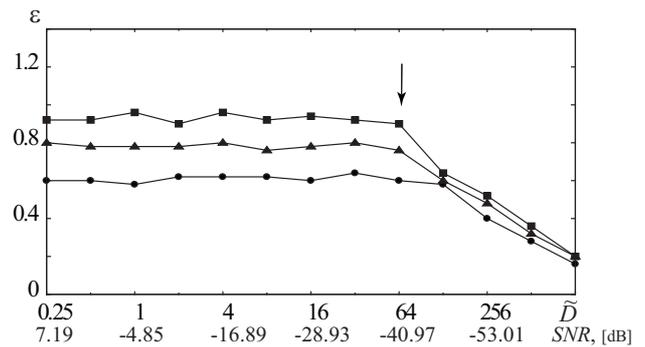}}}\caption{Dependence
of the boundary value of the GS regime onset in the coupled CGLE's
on the noise intensity $\tilde D$ (the SNR values corresponding to
the noise intensities are also shown) for different values of the
control parameters of the response system: $\alpha_r=3, \beta_r=3$
({\Large $\bullet$}), $\alpha_r=4, \beta_r=4$ ({\large
$\blacktriangle$}), $\alpha_r=5, \beta_r=5$ ($\blacksquare$). The
critical value of the noise intensity $D_c$ up to which the
\col{boundary value of the GS regime does not almost depend on the noise intensity}
is marked by arrow \label{fgr:EGS_noise}}
\end{figure}

\begin{figure*}[tb]
\centerline{\scalebox{0.7}{\includegraphics{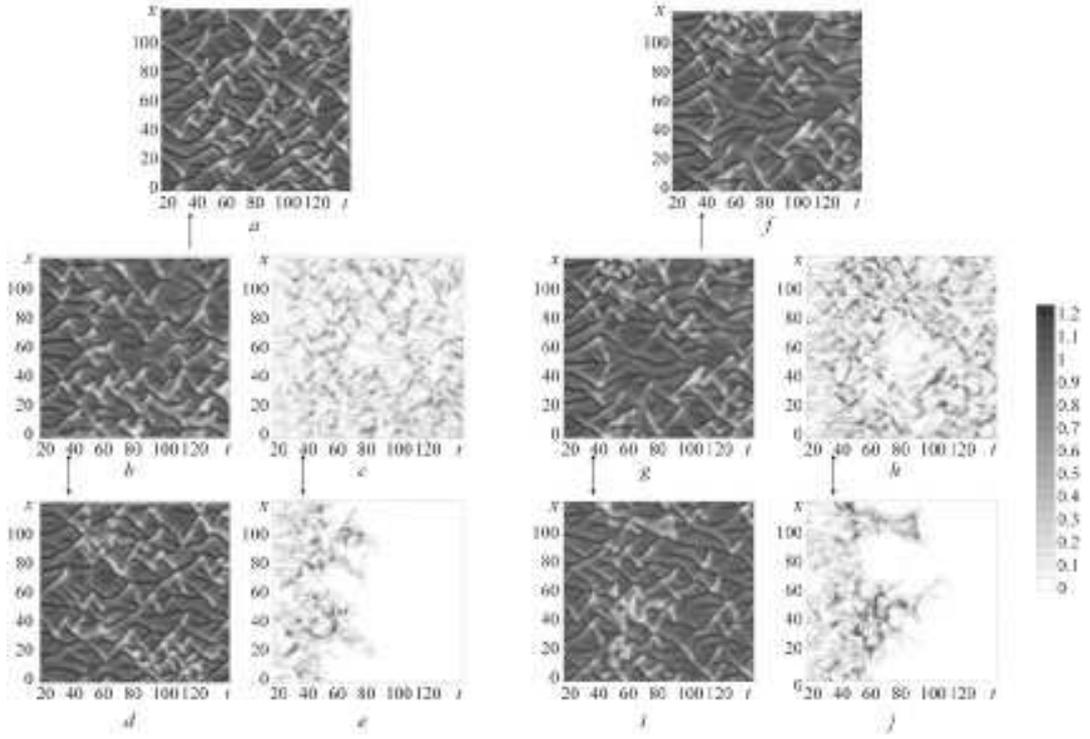}}} \caption{The
spatio-temporal diagrams characterizing behavior of the drive
(\textit{a,f}) and response (\textit{b,d,g,i}) systems
(\ref{eq:CGLE_drive})--(\ref{eq:CGLE_response}) as well as the
dependencies of the module of the difference between the states of
the response and auxiliary systems  $|v-v_a|$ (\textit{c,e,h,j}) for
cases of absence (\textit{c,h}) ($\varepsilon=0.2$) and presence
(\textit{e,j}) ($\varepsilon=0.8$) of the GS on time $t$ and space
$x$. The control parameter values for the response system have been
selected as $\alpha_r=\beta_r=3$. The time moments marked by arrows
correspond to the coupling switching-on between the drive and
response systems. Pictures (\textit{a--e}) correspond to the
noiseless case ($\tilde{D}=0$) whereas (\textit{f--j}) refer to the
noise one ($\tilde{D}=0.4$) \label{fgr:TimeSpaceGL}}
\end{figure*}

Fig.~\ref{fgr:EGS_noise} shows the dependence of the boundary value
of the GS regime onset $\varepsilon$ on the noise intensity $\tilde
D$ (SNR value) for several values of the control parameters of the
response system.
One can easily see that the noise of intensity ${\tilde D \in
[0;64]}$ ($\rm SNR\geq -41dB$) does not almost \col{affect the
threshold} of the GS regime onset in spatially extended systems
described by the Ginzburg-Landau equations. The time-space diagrams
characterizing the behavior of unidirectionally coupled spatially
extended media before and after the GS regime onset both in the
absence and presence of noise are shown in
Fig.~\ref{fgr:TimeSpaceGL}.
Pictures (\textit{a--e}) correspond to the case of the absence of
noise both in the drive and response systems whereas in the pictures
(\textit{f--j}) the white noise of intensity $\tilde{D}=0.4$ affects
both the drive and response. Pictures (\textit{a,f}) characterize
the drive system behavior, whereas the other ones refer to the
response system one before (\textit{b,g}) and after (\textit{d,i})
the GS regime onset. Fig.~\ref{fgr:TimeSpaceGL},\textit{c,e,h,j}
shows the spatiotemporal distributions of the module of the
difference between the states of the response and auxiliary systems
$|v-v_a|$ for cases of the absence (\textit{c,h}) and the presence
(\textit{e,j}) of the GS regime. One can easily see, that in the
second cases (\textit{e,j}) the difference of the states of the
response and auxiliary systems in every point of space tends to be
zero after coupling begins, which means the presence of the GS
between the drive and response CGLE's. It should be noted that the
length of the transient process preceded to the GS regime onset is
occurred to be rather more in the case of the presence of noise
whereas the threshold value of the GS regime onset is the same as in
the noiseless case. Moreover, one can easily see that
spatio-temporal diagrams characterizing the response system behavior
are similar to each other both in the presence and absence of noise
(compare pictures (\textit{b,d}) with (\textit{g,i}), respectively).

The stability of the GS regime in Ginzburg-Landau equations
\col{with respect to} noise is determined by the same mechanisms, as
in the cases of the systems with a few number of degrees of freedom
considered in the previous subsections~\ref{sct:LogMaps} and
\ref{sct:Roessler}. As well as for the logistic maps and R\"ossler
systems, the modified system approach may be used for the
explanation of the observed phenomenon. Indeed, the noise of a large
enough intensity does not almost \col{affect the characteristics} of
the modified Ginzburg-Landau equation
\begin{equation}
\begin{array}{l}
\displaystyle \frac{\partial v_m}{\partial t}=
v_m-(1-i\alpha_r)|v_m|^2 v_m+ \\ \displaystyle \quad
+(1+i\beta_r)\frac{\partial^2 v_m}{\partial x^2} - \varepsilon v_m,
\quad x\in[0,L]
\end{array}
\label{eq:modifiedCGLE}
\end{equation}
(and, as a consequence, of the response one), as well as in the case
of Ginzburg-Landau equation with the added constant
term~\cite{Hramov:2008_INIS_PRE}. Therefore, the noise does not
change the threshold value of the GS regime onset. At the same time,
as it has been discussed in Section~\ref{sct:GS}, the boundary value
of the coupling parameter $\varepsilon$ may start changing if the
noise intensity is a very great (${\tilde{D}>64}$, ${\rm
SNR<-41dB}$). It is easy to see from Fig.~\ref{fgr:EGS_noise} that
for such values of the noise intensity the boundary value of the GS
regime starts decreasing. For the very large intensities $D$ the
coupling value $\varepsilon_{GS}$ corresponding to the boundary of
the GS regime tends to the constant value which does not depend on
the the control parameters $\alpha$ and $\beta$ of the spatially
extended media. Such behavior of the boundary value of the GS regime
onset, as in the case of unidirectionally coupled R\"ossler systems
considered in Section~\ref{sct:Roessler}, is connected with the
noise-induced synchronization regime realization.

Nevertheless, the noise of a large enough intensity does not almost
alter the threshold value of the coupling parameter strength between
two unidirectionally coupled Ginzburg-Landau equations. In this case
one can say about stability of the GS regime \col{with respect} to
noise in the coupled spatially extended self-oscillating media.

So, having considered three different examples of model systems
(discrete maps, flow systems, spatially-extended media) we can come
to the conclusion that the GS regime demonstrates the significant
stability \col{with respect} to noise in a wide range of the values
of the external noise intensity.

\section{Experimental study of the GS onset in chaotic circuits in the presence of noise}
\label{sct:Experiment} To confirm the theoretical and numerical
results given in the previous sections we have also studied
experimentally the dynamics of the chaotic oscillator driven by the
external chaotic signal in the presence of noise. In the experiment
we have used the simple electronic circuit where all parameters
including noise amplitude may be controlled precisely.

\begin{figure}[tb]
\centerline{\scalebox{0.45}{\includegraphics{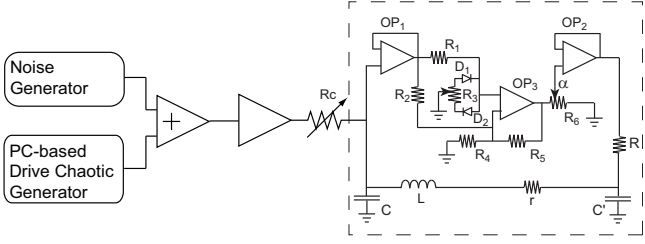}}}
\caption{Block diagram of experimental setup. Chaotic generator
layout is shown by dashed rectangle. Here $C$=330~nF,
$C^\prime$=150~nF, $R$=630~$\Omega$, $r$=56~$\Omega$, $L$=3.3~mH,
$OP_{1,2}$ -- TL082, $OP_{3}$ -- LF356N, $D_{1,2}$ -- 1N4148,
$R_1$=2.7~$k\Omega$, $R_2$=$R_4$=7.4~$k\Omega$, $R_3$=100~$\Omega$,
$R_5$=186~$k\Omega$, $R_6$=4.7~$k\Omega$, $RC^\prime$, $rLC$ --
low-pass filters, $\alpha$ is parameter of nonlinear converter,
$R_c$ is a coupling resistance \label{pct:ExpSetup}}
\end{figure}

\begin{figure}[tb]
\centerline{\scalebox{0.4}{\includegraphics{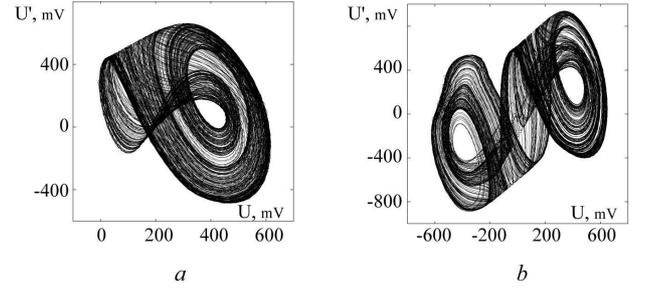}}}
\caption{Typical phase portraits of chaotic regimes observed in
experiment: (a) band attractor ($\alpha\sim 0.15$), (b) double
scroll attractor ($\alpha\sim 0.25$) \label{pct:CharRegimes}}
\end{figure}

The experimental setup is shown in Fig.~\ref{pct:ExpSetup}. As a
basic element of the scheme we have used an electronic circuit with
nonlinear converter and linear feedback loop similar to the one
described
in~\cite{Rulkov:1996_SynchroCircuits,Hramov:2007_TypeIAndNoise} (it
is shown in Fig.~\ref{pct:ExpSetup} by dashed rectangle). Since the
generator is capable to demonstrate both periodic and chaotic
oscillations depending on the choice of the parameter $\alpha$ of
nonlinear converter, it has been selected in such a way for the
generated signal to be chaotic (quantitative values of all control
parameters of the circuit are presented in the captions of
Fig.~\ref{pct:ExpSetup} and \ref{pct:CharRegimes}). Chaotic
generator has been connected to DAC/ADC board L-Card L-783 installed
into personal computer (PC) whereby we have recorded the dynamics of
potential on the capacitors $C$ and $C^\prime$.
As a drive signal we have used the last one generated by the circuit
described above, digitized by ADC with further reconstruction by
DAC. The drive signal has been introduced into the circuit via
dissipative unidirectional coupling of variable dissipation value
(see Fig.~\ref{pct:ExpSetup}). The noise signal has been produced
with Agilent 33220 function generator, digitized and additively
introduced into the coupling device (as it has been shown in
Fig.~\ref{pct:ExpSetup}). Characteristics of the noise are close to
the Gaussian one. Oscillations of the response system have been also
digitized with ADC board and transferred to personal computer for
further numerical processing.

As we have mentioned above, one of the easiest ways to detect the
presence of the GS regime is the use of an auxiliary system, i.e. an
additional response circuit, which is a replica of the main one. But
creation of the auxiliary system with parameters completely equal to
the response system ones is one of the most conceptual problems in
the experimental study of the GS regime. To solve this problem we
have used an approach analogous to the one discussed
in~\cite{Uchida:2003_GSLaserPRL}. As it has been specified above,
the signal from the drive system with additive noise has been
preliminary recorded on PC. Therefore, it is evident that in this
case the response system could be subjected to the influence of
identical drive signal (with \col{additive} noise) any number of
times. For the realization of an auxiliary system method it is quite
sufficient to \col{affect the response system} by the drive signal
twice, alternating the period of the influence with the time
interval of autonomous dynamics (to provide the different initial
conditions), and then compare obtained data numerically.

The experiment has been performed for three main cases: (i) chaotic
attractors both in the drive and response systems have identical
band structure;  (ii) chaotic regime with band attractor has
influenced on the regime with double scroll attractor; (iii) chaotic
attractors both in the drive and response systems have a double
scroll structure. Typical phase portraits of considered regimes are
shown in Fig.\ref{pct:CharRegimes} (band attractor (a) and double
scroll attractor (b)). The corresponding values of the control
parameter $\alpha$ are indicated in the caption.
Each case has been studied in the presence of Gaussian noise of
different intensity. For experimental data the noise intensity has
been calculated as a ratio ${D=P_N/P_{CS}}$ of a power of the noise
signal $P_N$ to the power of chaotic signal $P_{CS}$.

\begin{figure}[t]
\centerline{\scalebox{0.45}{\includegraphics{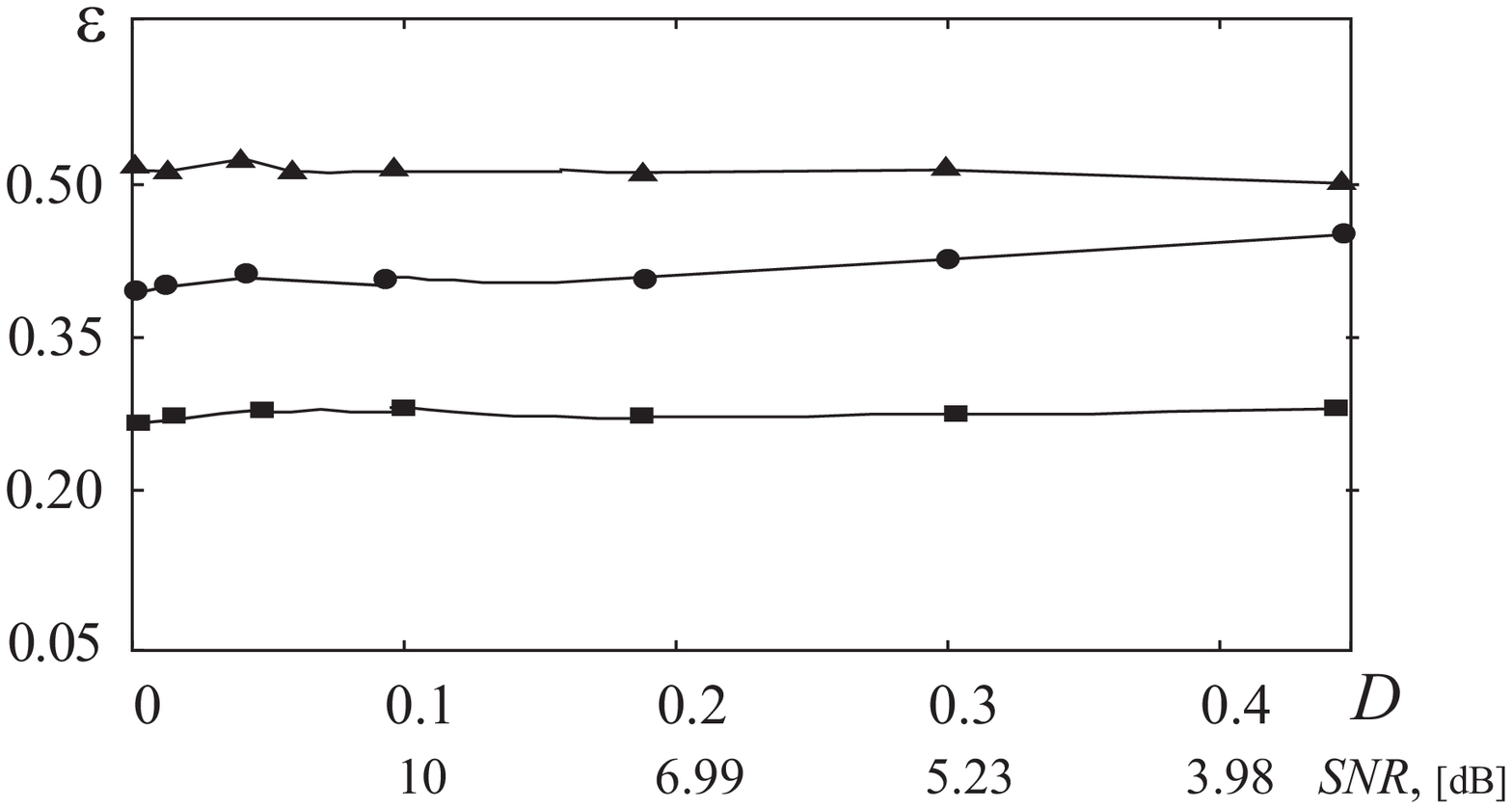}}}
\caption{Coupling strength value corresponding to the GS regime
onset as a function of noise intensity (the SNR values corresponding
to the noise intensities are also shown) in the cases when chaotic
attractors both in drive and response systems have the band
structure ($\blacksquare$); drive system in the band chaotic regime
influences on the response system in chaotic regime with the double
scroll attractor ({\Large $\bullet$}); chaotic attractors both in
drive and response systems have a double scroll structure ({\Large
$\blacktriangle$}) \label{pct:Results}}
\end{figure}

\begin{figure}[tb]
\centerline{\scalebox{0.62}{\includegraphics{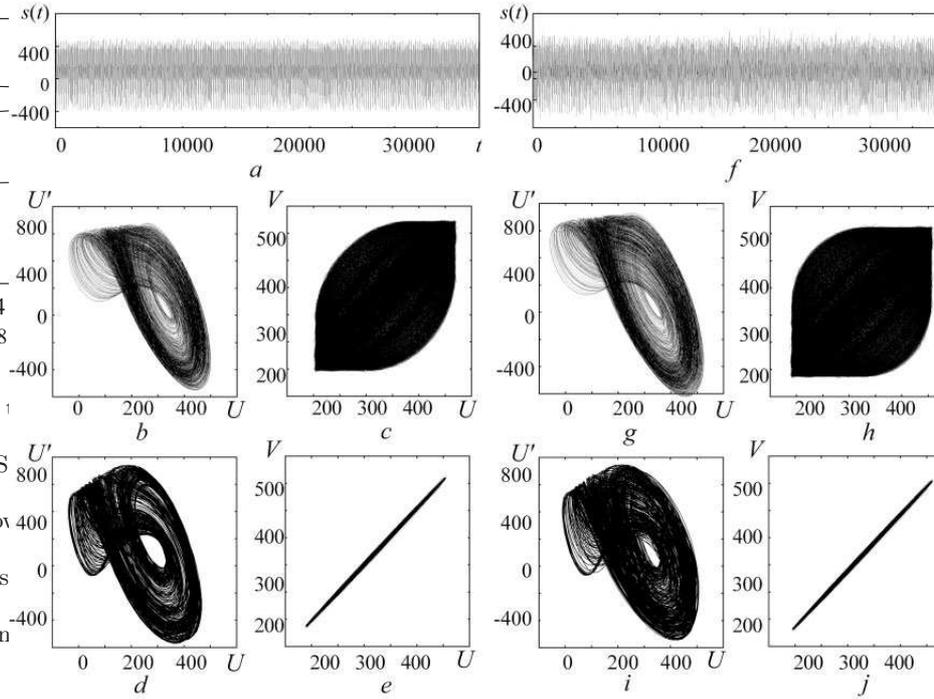}}}
\caption{Signals from the drive chaotic circuit without (\textit{a})
and with (\textit{f}) \col{additive} noise \col{affecting the response}
circuit, phase portraits of the response system (\textit{b,d,g,i})
and $(U,V)$-planes characterizing the response and auxiliary system
behavior (\textit{c,e,h,j}) before ($\varepsilon=0.22$) and after
($\varepsilon=0.34$) the GS regime onset. The control parameters of
the chaotic circuit has been chosen in such a way that both the
drive and response systems in autonomous regime are characterized by
the band attractors. Pictures (\textit{a--e}) correspond to the
noiseless case ($D=0$) whereas (\textit{f--j}) refer to the noise
one ($D=0.4$) \label{fgr:PhasePortRulkov}}
\end{figure}

Figure~\ref{pct:Results} shows the dependence of the coupling
strength value $\displaystyle
\varepsilon=\frac{1}{R_c}\sqrt{\frac{L}{C}}$ corresponding to GS
regime onset on the noise intensity for three cases mentioned above.
One can see that in the range of noise intensity [0;~0.5] the
threshold value remains nearly constant. Typical signals from the
drive system with and without \col{additive} noise \col{affecting
the response} one as well as the phase portraits of the response
system and $(U,V)$-planes characterizing the response and auxiliary
system behavior both in the absence and presence of the GS regime in
chaotic circuits in the case (i) are shown in
Fig.~\ref{fgr:PhasePortRulkov}.
One can easily see that characteristics of the response circuit have
not been changed noticeably with the appearance of noise. Analogous
situation takes place in unidirectionally coupled chaotic circuits
with initially double-scroll chaotic attractors in the one and both
of them. One can say that in all considered cases the modified
system (i.e. considered generator with additional dissipation)
demonstrates the cycle-1 periodic oscillations. Further increase of
the noise intensity (when it becomes greater than the intensity of
the deterministic signal) may result in the monotonous growth of the
GS boundary value.

So, the experimental results satisfy the stability of the GS regime
\col{with respect} to noise. They are also in a good agreement with
the data obtained theoretically and numerically.

\section*{Conclusions}
\label{sct:Conclusions} In conclusion, we have analyzed both
theoretically and experimentally the influence of noise on the GS
regime in different unidirectionally dissipatively coupled identical
chaotic systems with mismatched parameters with a small number of
degrees of freedom as well as spatially extended media. The
dependencies of the GS regime boundary on the noise intensity in the
cases when the drive and response systems are enforced both by
common noise and by the statistically independent noise sources are
also considered. We have shown that if attractors of the drive and
response systems have an \col{infinitely large} basin of attraction,
independently on the type of system and kind of the noise
distribution the GS regime possess a great stability \col{with
respect} to noise, i.e. the threshold of the synchronous regime
arising does not almost depend on the intensity of noise. Such
behavior of the boundary of the GS regime has been explained by
means of the modified system approach, i.e., the joint action of
dissipation and driving force is responsible for the reported
robustness of the GS regime against noise.

Though the results described in the Manuscript refer to the white
noise we expect that they could be valid for different noise forms.
Similar results have been obtained for different types of noise,
including colored noise.

It should be noted that the revealed peculiarity of the GS regime
could be used in a number of practical applications, i.e. for the
transmission of information through the communication channels where
the level of noise is
sufficient~\cite{alkor:2010_SecureCommunicationUFNeng,Moskalenko:InfoTransNoisePLA2010}.


\end{document}